\documentstyle[12pt,epsfig,epsf]{article}

\newcommand{\institute}[1]{\parbox{16cm}{%
\centering\normalsize \sl #1}}
\def\lsi{\raise0.3ex\hbox{$<$\kern-0.75em\raise-1.1ex\hbox{$\sim$}}}
\def\gsi{\raise0.3ex\hbox{$>$\kern-0.75em\raise-1.1ex\hbox{$\sim$}}}

\title{Better than \$1/Mflops sustained: a scalable PC-based parallel computer for 
lattice QCD}
\author{%
Zolt\'an Fodor$^1$\thanks{email: fodor@poe.elte.hu}, 
S\'andor D. Katz$^2$\thanks{on leave from
Institute for Theoretical Physics, E\"otv\"os University,
P\'azm\'any P. 1/A, H-1117 Budapest, Hungary, \hspace*{6mm}email: sandor.katz@desy.de},  
G\'abor Papp$^1$\thanks{email: pg@ludens.elte.hu}\\
\institute{
$^{(1)}$ Institute for Theoretical Physics, E\"otv\"os University,
P\'azm\'any P. 1/A, H-1117 Budapest, Hungary,\\
$^{(2)}$ Deutsches Elektronen-Synchrotron DESY, Notkestr. 85, D-22607,
Hamburg, Germany}
}
\date{11 January, 2002}

\begin{document}
\psfull

\maketitle

\vspace{-8.5cm}
{ \normalsize
\hfill \parbox[t]{4cm}{ITP-Budapest 580 \\ DESY 02-018\\
                  } \\[7em]
\vspace{6.2cm}

\begin{abstract}
We study the feasibility of a PC-based parallel computer
for medium to large scale lattice QCD simulations. 
The E\"otv\"os Univ., Inst. Theor. Phys. cluster consists
of 137 Intel P4-1.7GHz nodes with 512 MB RDRAM. The 32-bit,
single precision sustained performance for dynamical QCD
without communication is  1510 Mflops/node with Wilson 
and 970 Mflops/node with staggered fermions.
This gives a total performance
of 208 Gflops for Wilson and 133 Gflops for staggered QCD, respectively
(for 64-bit applications the performance is approximately halved).
The novel feature of our system is its
communication architecture. In order to have a
scalable, cost-effective machine we use Gigabit Ethernet cards for
nearest-neighbor
communications in a two-dimensional mesh. 
This type of communication is cost effective
(only 30\% of the hardware costs is spent on the communication). According to 
our benchmark measurements this type of communication
results in around 40\% communication time fraction for 
lattices upto $48^3\cdot96$ in full QCD simulations. 
The price/sustained-performance ratio for full
QCD is better than \$1/Mflops for Wilson (and around \$1.5/Mflops for 
staggered) quarks 
for practically any lattice size, which can fit in our parallel computer. 
The communication software is freely available upon request for non-profit 
organizations.
\end{abstract}

\section{Introduction}

The most powerful computers for lattice gauge theory are industrial
supercomputers or special purpose parallel computers (see e.g.
\cite{APE,APE-project,Tsukuba,Columbia}).   
Nevertheless, it is more and more accepted that even off-the-shelf processors 
(e.g. PC processors) can be used to build 
parallel computers for lattice QCD simulations 
\cite{Wuppertal,Jlab-MIT,CSea99,Lea01}.  

There are obvious advantages of PC based systems. 
Single PC hardware usually has excellent price/performance ratios 
for both single 
and double precision applications. In most cases 
the operating system (Linux), compiler (gcc) and other software are 
free. 
Another advantage of using PC/Linux based systems is that lattice codes 
remain portable. Furthermore, due to their price they are available for 
a broader community working on lattice gauge theory. 
For recent review papers and benchmarks see 
\cite{C99,G00,L01}.

Despite the obvious advantages it is not clear how good is the 
price/performance ratio in medium or large scale lattice QCD applications,
for which a large number of processors are needed. Clearly, these
multi-processor computer systems dissipate a non-negligible amount of heat.
The components should be reliable in order to avoid frequent system crashes.
Most importantly, the communication network should be fast and cheap enough
in order to keep the good price/performance ratio of the single PCs.

We present our experiences and benchmark results. We show that a parallel 
system e.g. based on ${\cal O}(100)$ PCs can run for several weeks or 
more without 
system crashes. The costs for air-conditioning are quite small. 
Such a system is scalable, it uses nearest-neighbor
communication through Gigabit Ethernet cards. 
The communication is fast enough (consuming 40\% of the total time
in typical applications) and cheap enough (30\% of the total 
price is spent on the communication).   
The system can sustain 100-200 Gflops on today's medium 
to large lattices.  This result gives a price/performance ratio
better than \$1/Mflops for 32-bit applications (twice as much for 
64-bit applications). 

Already in 1999 a report was presented \cite{CSea99}
on the PC-based parallel computer project at the
E\"otv\"os University, Budapest, Hungary. 
A machine was constructed with 32 PCs arranged in a three-dimensional 
2$\times$4$\times$4 mesh. Each node had two special, purpose designed
communication cards providing communication through flat cables 
to the six neighbors. We used the ``multimedia extension'' instruction set
of the processors to increase the performance.
The bandwidth between nodes (16 Mbit/s) was 
sufficient for physical problems, which needed large bosonic systems
\footnote{SU(2)-Higgs model for the
finite temperature electroweak phase transition (for simulation techniques see
e.g. \cite{ewpt})}; 
however, the communication was far 
too slow for lattice QCD. Here we report on a system which needs
no hardware development and has two orders of 
magnitude larger communication bandwidth through Gigabit Ethernet cards.
Our system is based on PCs, each with a P4-1.7GHz processor and
512 MB RDRAM memory. This system with its Gigabit Ethernet communication
represents a good balance between CPU power and network bandwidth. According
to our knowledge this is the first scalable parallel computer for lattice
QCD with a price/sustained-performance ratio better than \$1/Mflops.

In Section 2 we briefly describe the hardware and discuss its cost.
Section 3 deals with the software used by our system, particularly
with the C library we developed for communication. In Section 4 we 
discuss physics issues and present benchmark
results based on the MILC code in Section 5. Section 6
contains our conclusions.

\section{Hardware}
The nodes of our system are almost complete PCs 
(namely PCs without video cards, floppy- and CD-drives)
with four additional 
Gigabit Ethernet cards.
Each node consists
of an 
Intel KD850GB motherboard,
Intel-P4-1.7GHz processor,
512~MB RDRAM,
100~Mbit Ethernet card,
20.4~GB IDE HDD and four SMC9452 Gigabit Ethernet cards
for the two dimensional communication.  

\begin{figure}[t]
\epsfig{file=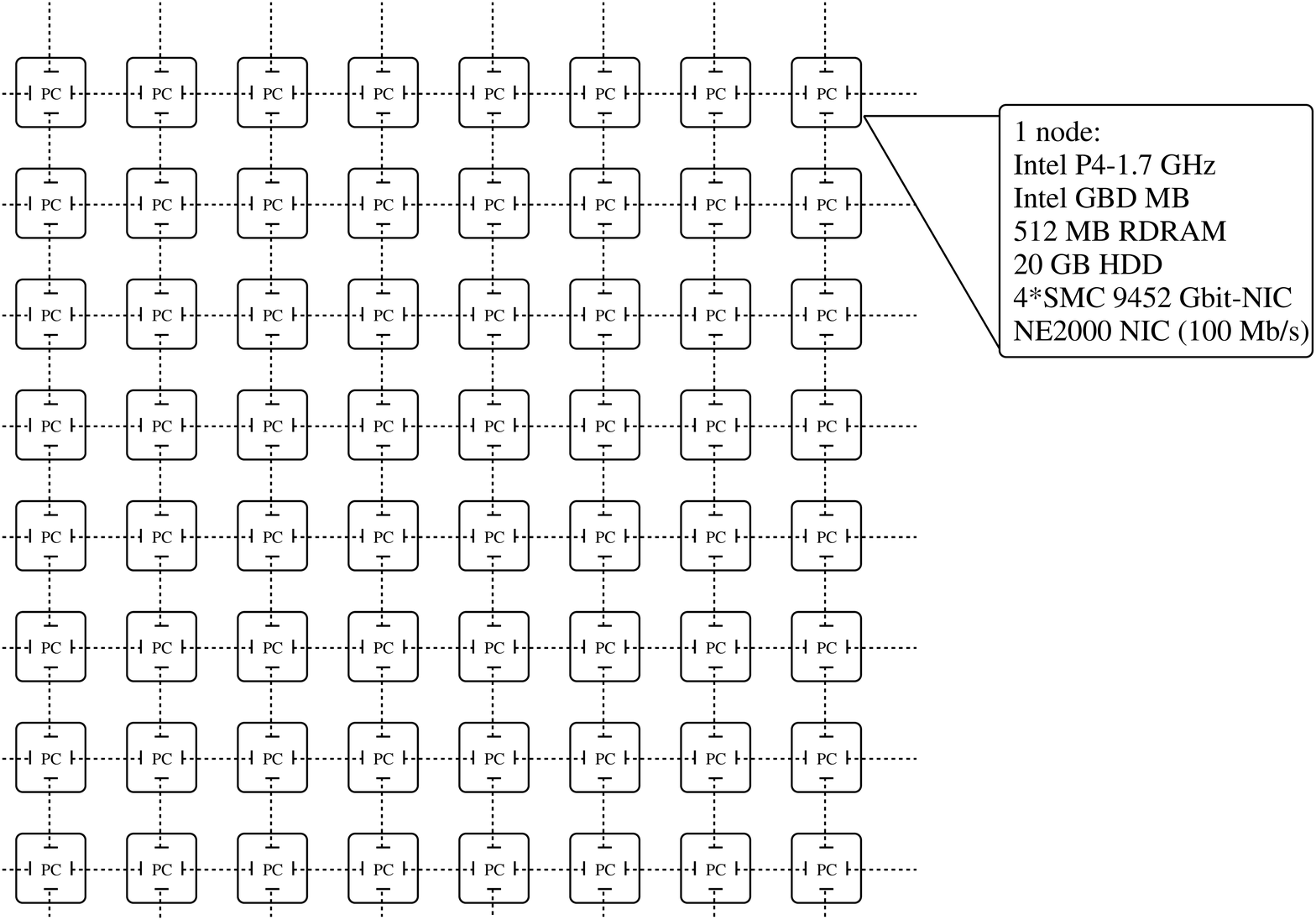,height=12cm}
\caption[hl]{Illustration of a 64-node cluster. The PC nodes
are arranged in a 8$\times$8 two-dimensional torus. Each node is an almost
complete PC with four additional Gigabit Ethernet cards. The four
Gigabit cards are used to reach the four neighboring nodes through
cross-twisted copper Cat5 UTP cables. 
As indicated,
at the boundaries periodic boundary conditions are realized.
\label{rms}}
\end{figure}

Altogether we have 137 PC nodes. One might use it as one cluster with 
128 nodes (or two clusters with 64 nodes or four clusters with 32 nodes) 
for mass production\footnote{We have 264 SMC9452 cards and our benchmark 
results were obtained on 64-, 32-, 16-, 8- and 4-node clusters}. 
Two smaller clusters with 4 nodes are installed 
for development with or without communication.
One additional, single machine is used for compilation, controlling 
the system and for smaller tests. 

After turning the machines on
each node boots Linux from its own hard-disk. 
All nodes can be accessed through a 100~Mbit Ethernet network with switches. 
The single node  (the ``main'' computer) controls the whole cluster.
A simple job-management package has
been written to copy the executable
code and data to and from the nodes and to execute the programs.

Each motherboard has 
five
PCI slots (32bit/33MHz). One of them is used
for the 100Mbit Ethernet card\footnote{note, that another version of
this motherboard is equipped with the 100~Mbit Ethernet
interface, thus all five PCI slots can be used for other purposes}.
These cards are connected to six Ovislink FSH24TX+ 24 port 100~Mbit
Ethernet switches. Controlling and copying files from and to the
nodes  
is done through this network.
The other PCI slots
contain the Gigabit Ethernet cards.  Each Gigabit Ethernet card is
connected to a Gigabit card of a neighboring node by a
cross-twisted copper Cat5 UTP cable. This allows us to perform
nearest-neighbor 
communication.  The number of the PCI slots on the
motherboard determines the dimensionality of the resulting mesh. In
our case there are five slots. Four
slots are used for Gigabit communication in four directions, thus in two
dimensions (see Figure~\ref{rms}). In principle, we could have  used
the remaining fifth PCI slot for a third dimension of the mesh, though
in this third dimension the mesh would have only two nodes  (due to
the periodic boundary condition a single connection is enough in this
special case). This arrangement results in a 2$\times$8$\times$8 mesh
for 128 nodes. Similarly, other PCI slots can also be used for
dimensions with extension of 2 nodes. This idea could result in meshes
e.g. 2$\times$2$\times$2$\times$8  or 2$\times$2$\times$8  for five
or four PCI slots with Gigabit connections, respectively\footnote{note, that 
having 2 nodes in one dimension does not reduces the communication bandwidth,
which is essentially fixed by the PCI bus speed}. The best
node-topology for a given lattice can be determined by optimizing the
surface to volume ratio of the local sub-lattice. 
Changing the node topology needs reconnecting the
Cat5 UTP cables of the system. This procedure is extremely easy, 
one person can do it in less
than 10 minutes. Note, that presently even 32 or 64 nodes can
accommodate fairly large lattices.  The topology of e.g. a  64-node
system is 8$\times$8 with periodic boundary conditions.  Such a
cluster is illustrated on Figure~\ref{rms}.  Note, that the
architecture of the machine is similar to that of the APE 
machines~\cite{APE}.

We started to use our cluster in October 2001. In this three months we never
suffered from a system crash of a running PC due to hardware or OS problems. 
According to our experiences the system can run for several weeks 
without needing a reboot (after which we reboot anyhow). 

One of the most important advantage of PC-based parallel computers is their
exceptional price (for price estimates see e.g. \cite{pricewatch}, though 
for ${\cal {O}}$(100) PCs one can easily reach better prices). 
In February, 2002 the price of one node including the 100~Mbit Ethernet 
switches is \$502. The four Gigabit Ethernet cards 
cost additional \$240 for each node 
(including the cables, which is less than \$1 for
each).  The power consumption of one node is 140W.
Since the machines dissipate a non-negligible amount of heat, we
had to install a cooling system. The costs of the cooling system, when
distributed among the individual nodes, resulted in \$13 for one node.
Thus, the total node (PC+communication+cooling) price is \$755.

As we will discuss it in Section 4 the typical performance of one node with
communication overhead is around 800-900~Mflops for QCD with Wilson 
(500-600~Mflops with staggered) quarks. Note, that whenever we speak about
performance we mean sustained-performance.  Comparing these performance 
results and the above prices we end up with a price/performance ratio 
better than \$1/Mflops for QCD with Wilson 
(and around \$1.5/Mflops with staggered) quarks.

There is an important message about our architecture.
The key element of a cost effective design is an appropriate 
balance between communication and the performance of the nodes.
As it can be seen we spent more than twice as much on the bare PC (\$515
including cooling) than for the fast Gigabit communication (\$240). 
These numbers are in strong contrast with
Myrinet based PC systems, for which the high price of the Myrinet card
exceeds the price of such a PC by a factor of two~\cite{myricom}. Similarly,
motherboards with 64bit/66MHz PCI buses are expensive, too.
Since even for our Gigabit Ethernet based machine the communication
overhead is only around 40\% (and in principle could be halved) 
a faster communication can not speed up significantly the system.
Clearly, these ratios result in a far worse price/performance relationship
for Myrinet based systems than for our architecture. 

As we will see the time needed for the calculation is dominated by memory
access. 
This can not be made better by having faster 
communication or processor.
There are in principle two ways to improve the
obtained price/performance ratio for PC-based systems.
First of all, one can push the costs of such a system down. 
This is realized in
our case by choosing relatively cheap PC components with the high memory 
bandwidth RDRAM and a rather cheap communication. Secondly, one might use
much faster memory access for a comparable price  
(e.g. use even faster memory access on the motherboard
or put the whole lattice to the cache memory). In this case the faster
communication would pay off. Unfortunately, this second option can not be
used cost effectively right now. 

\section{Software}

The main operating system of the cluster is SuSE Linux 7.1, being
installed on each node, separately. As it was mentioned earlier, the 
machines are connected via Gigabit Ethernet cards, 
4 cards per machine building
up a 2 dimensional periodic matrix with 4 neighbors for each node (with
computers on the side being connected to the one on the opposite side).
The job management is done by the ``main'' computer through an additional 
100~Mbit Ethernet network (see e.g.  \cite{CSea99}). This separate
PC is where the main part of the program development and compilation is done
and only the binary code is copied to the nodes. Communication development 
is done on a separate small 4-PC cluster.

Nodes have special hostnames indicating their location in the matrix.
For the 128-node cluster the names are {\tt p00,p01 $\dots$ p07,p10,p11 
$\dots$ pf7}. The two hexadecimal
digits after the leading p represent the two coordinates of the node
within the 16$\times$8 matrix.

In order to take advantage of the Gigabit communication from applications
(e.g. C, C++ or Fortran code), a simple C library was written using the
standard Linux network interface. Currently, we are using a standard socket
based communication with Transmission Control Protocol (TCP) widely used
in network applications. Clearly, TCP is much more advanced than our needs 
for a nearest-neighbor
communication. This results in a somewhat large 
communication overhead. As we discuss it later, the typical
measured communication bandwidth in QCD applications is around 400 Mbit/s,
which should be compared with the maximal theoretical bandwidth of
1000 Mbit/s and with the 800 Mbit/sec measured bandwidth for large packages. 
In order to eliminate the overhead due to TCP/IP we are
working on a new, more efficient driver.

During the startup period of the connection
odd nodes act as clients, while even nodes are the servers (for the node
with hostname {\tt pxy} the parity is defined by the parity of
{\tt x+y}). After 
the network sockets (one two-way socket per link) are established they are 
kept open till the  end of the program. During the execution of the program 
the previous ``clients'' and 
``servers'' act symmetrically,  both sending and receiving information 
to or from the neighboring nodes.

The functions of the library handle the data exchange between the
neighbors. They can be summarized as follows (the prefix ``gb'' refers to
Gigabit).

\vspace{.5cm}
\noindent\verb/int gb_open(int *nodes_x, int *nodes_y)/

Opens four two-way network sockets to the neighboring nodes. On each
node hostnames {\it dir0} (right), {\it dir1} (up), {\it dir2} (down) and
 {\it dir3} (left) are defined to the IP address of the appropriate
neighboring Ethernet card in the {\it /etc/hosts} file. 
First, the number of available machines is checked in the two directions and
the variables {\it nodes\_x} (left-right) and {\it nodes\_y} (up-down)
are set accordingly. 
On success {\it gb\_open} returns NULL, otherwise -1.

\vspace{.5cm}
\noindent\verb/int gb_close()/

Closes all four network sockets.
No further communication is possible after this call.
On success {\it gb\_close} returns NULL, otherwise -1.

\vspace{.5cm}
\noindent\verb/int gb_send(int dir, char *buffer, int size)/

\noindent\verb/int gb_recv(int dir, char *buffer, int size)/

Sends/receives data to/from the given direction. Directions are 
0 (right), 1 (up), 2 (down) and 3 (left). Data are provided through
a pointer {\it buffer} with length {\it size} in bytes. In case of
receive operation {\it size} is the length of the buffer provided.
On success the number of transferred bytes is returned, otherwise
the return value is -1.

\vspace{.5cm}
\noindent\verb/int gb_send_recv(int dir, char *send_buffer, char
*recv_buffer
int size)/

A combination of send and receive. Sends data to the given direction
and receives from the opposite direction. {\it send\_buffer} and
{\it recv\_buffer} are pointers to the data to be sent/received.
On success the total number of transferred bytes is returned (2 $\times$
{\it size}), otherwise the return value is -1. This function expects
exactly {\it size} bytes of data to be received and blocks other
operation until the whole amount arrives.

\vspace{.5cm}
\noindent\verb/int gb_collect(int dir, char *buffer, char *result, int size)/

Collects data from all the nodes in the direction {\it dir}
to the array {\it result}. The buffer size should be at least the number
of nodes in the given direction multiplied by {\it size}. The 
data to be transmitted from the current node is placed in {\it buffer}. 
It will be copied to the first {\it size} bytes in {\it result} on that node. 
The data from other nodes will be packed after each other 
and ordered in {\it result} by the distance of the 
originating node in the given direction
{\it dir}.
On success the total number of transmitted bytes is returned.

Synchronization can be an important issue for parallel computing. For the
presented system it is solved by the architecture itself. 
Each node uses its own clock and waits until the neighbors provide the
necessary data to proceed in the program. Since the Gigabit Ethernet cards
have their own memory there is no need to synchronize the ``send'' and 
``receive''  instructions.

\section{Computational requirements for lattice problems}

Large scale lattice QCD simulations are focused on a few topics. They
might be divided into two computationally different classes.\\
{\bf a.} One of them is full QCD. In this case 
the most important feature of a parallel computer is its maximal CPU
performance in Gflops since most of the time is spent on the updating process. 
The computational requirements for new theoretical developments are similar 
to those of full QCD. \\
{\bf b.} The other class is heavy quark phenomenology (quenched or unquenched).
In this case not only the CPU performance, but also the
memory requirements for the 
measurements play a very important role. \footnote{
We carry out 
full QCD simulations (staggered quarks at finite temperature and/or
chemical potential \cite{FK01}), thus our presented computer with its memory is 
somewhat less appropriate for heavy quark phenomenology; however, its 
memory can be easily extended.}     

First we discuss full QCD.  Only a small fraction (12-18\% for large
volumes with Wilson fermions)  of our 137-node  system's total nominal
performance  --0.93 Tflops for single precision and 0.47 Tflops for
double precision  calculations-- can be obtained as a sustained value
for full QCD  simulations. Before we show the detailed benchmark
results it is  instructive to discuss some features of the
calculations.  The basic reason for the relatively small
sustained-performance/peak-performance ratio  is the memory bandwidth. 
If the pure computational performance is $P$, the memory bandwidth is $B$
and the number of accessed bytes per operations in the code is $F$ then the
maximal sustained performance we can reach (assuming that memory access
and computation is not performed simultaneously) is $P_s=1/(1/P+F/B)$.
The theoretical computational performance is $P=6.8$~Gflop/s since
using the P4 processor's {\tt sse} instruction set, 
four single precision operations can be performed in one clock cycle.
The maximum bandwidth of the RAMBUS architecture is $B=3.2$~GB/s.
The value of $F$ is 1.27~bytes/flop for Wilson and 1.81~bytes/flop for
staggered quarks (cf. \cite{G00}). 
This gives an upper limit for the sustained performances,
$\approx$1800~Mflops for Wilson and $\approx$1400~Mflops for staggered 
quarks. Note, that in both cases $P_s$ is dominated by the term F/B 
(essentially determined by the memory bandwidth) and not by 1/P (fixed by
the theoretical computational performance). 
With careful cache-management one can prefetch the next 
required data concurrently with calculation. By this technique one 
increases the
maximal performance to $P_s \approx B/F$. This estimate further emphasizes 
the importance of the memory bandwidth.

Now we discuss the question, how good is the presented PC based 
parallel computer for heavy quark phenomenology.
For this sort of physics problem a important computer issue
is the memory requirement for the measurement of a heavy-light form-factor.
At present we have 512 Mbytes/node. This gives for the
128 nodes 64 Gbyte, which is the memory of the possibly
largest APEmille system of 8$\times$8$\times$32 \cite{APE-project}). 
The memory of the presented system can be easily extended. One might
have 1.5 Gbyte/node, which increases the costs by approximately 35\%. 
This gives 200 Gbyte memory for a 128-node cluster. 
If only two propagators at the time are kept in memory, while the others 
are stored and reloaded from disks, memory requirements reduce sharply (see
e.g. the analysis of the apeNEXT project \cite{APE-project}). Note, that
this option needs a high bandwidth to and from the disks. Our measured 
total disk input/output bandwidth for a 128-node cluster is 1.2 Gbytes/s,
thus even very large propagators can be loaded or stored in less than a
minute. Due to the usage of the local disks this bandwidth scales with the 
number of nodes and the required time for input/output remains the same. 
Combining the above features it is easy to see that with
extended memory fairly large lattices, 
e.g. of the order of $64^3\cdot 96$ can be studied for heavy quark phenomenology 
on a 128-node cluster. 

\section{Performance}

\begin{figure}[t]\begin{center}
\epsfig{file=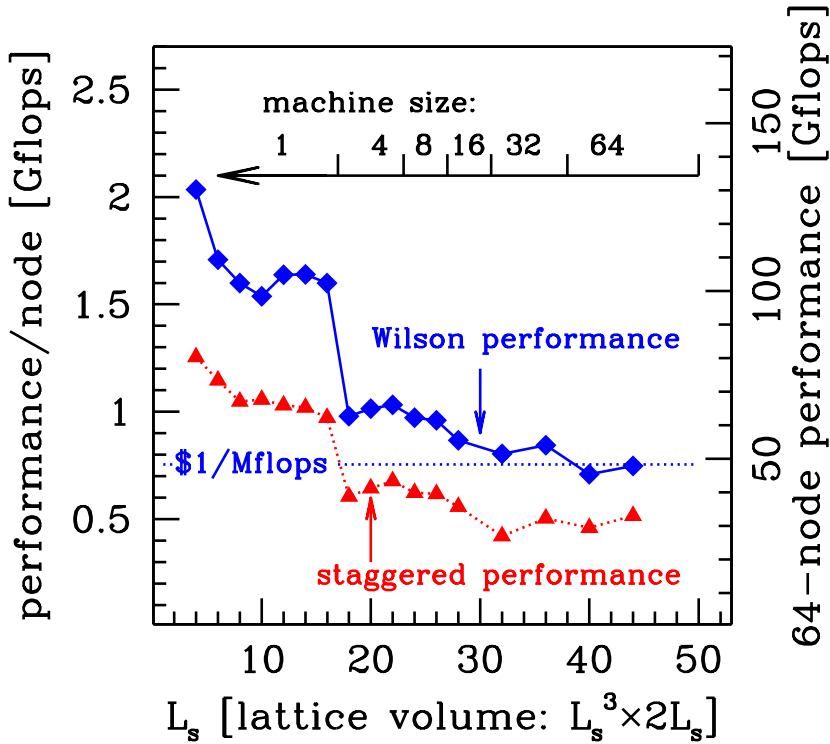,height=12.cm}
\caption[hl]{Full QCD performance results on a 64-node cluster.
We indicate the node performance (left vertical axis) and the total
performance of the 64-node cluster as a function of the spatial
extension $L_s$. The lattice sizes are chosen to be $L_s^3\times 2L_s$.
Squares indicate Wilson, triangles staggered dynamical QCD simulation 
performances, respectively.
The dotted line represents the \$1/Mflops value. 
Lattices of size $16^3\times$32 or smaller fits in the memory of one node. 
For these lattice
sizes single nodes were used without communication. For larger lattices
communication was switched on. This resulted in an $\approx 30$\% reduction
of the sustained performances.  
The inserted scale
shows the (effective) machine size as the number of
communicating nodes on which the lattice is distributed.
\label{perf}}
\end{center}\end{figure}

In this section we discuss the benchmark results for the full QCD 
case. Figure~\ref{perf} gives a summary of these runs. Single precision
is used for local variables (gauge links and vectors, see \cite{MILC}) 
and dot products are accumulated in double precision.
Using double precision gauge links and vectors reduces the performance by
approximately 50\%. The reason for this reduction is twofold. On the one hand, 
both the memory access, and the communication between the nodes needs twice 
as much time; on the other hand, sse2-performance (double precision) 
is just the half of the sse-performance (single precision).
In the simulations the R-algorithm is used \cite{G87}. Our
results correspond to the full code performances
(not only to the conjugate gradient iteration)
\footnote{For the staggered case the full code has been
tested using communication. For the Wilson case the computationally 
most important $(D_W+m_0)\psi$ is tested with communication. The full Wilson 
performance is estimated based on  $(D_W+m_0)\psi$, 
the same for staggered quarks ($(D_s+m_0)\psi$) and the total 
staggered performances.}.

For benchmarking we started from the MILC Collaboration's public code
~\cite{MILC}.
In order to increase the performance of the code we modified the
original MILC code by three different techniques.

First of all, we used the ``multimedia extension'' instruction set of the 
Intel-P4 processors (note, that the same sse instruction set is now available on
Athlon XP processors, too). 
As we pointed out it in 1999 \cite{CSea99} this capability
can accelerate the processor by a large factor (in Ref. \cite{CSea99}
we used AMD-K6 processors, which performs 4 operations for each clock cycle
by using its  multimedia extension). The multimedia extension of the
Intel-P4 processor works as a vector unit with 128 bits wide registers, which
can be used for arithmetic operations with 2 double precision or 4 single
precision numbers. There is a simple way to access these registers, work with
them and obtain very good performances (see M. L\"uscher's excellent review 
\cite{L01}). 
We rewrote almost the whole conjugate gradient part of the
program to assembly (including also loops over the lattice and not only
elementary matrix operations). No preconditioning was used.
The $SU(3)$ matrices and vectors were stored
in a format that fits well to the {\tt sse} architecture 
(see the appendix for details). 
This way we obtained
a speedup factor of $\approx$2 in the performances 
(similar benchmark results were obtained by e.g. Ref. \cite{P01}).  

Secondly, an important speedup was obtained by changing the data structure of
the original MILC code suggested by S. Gottlieb \cite{G01}. 
The original code is based on the ``site major'' concept.
This structure contains all the physical variables of a given site and the
lattice is an array of these structures.  Instead of this concept one should 
use ``field major'' variables.
The set of a given type of variable (e.g. gauge links) of the different
sites are collected and stored sequentially.
In this case the gauge fields and the 
vectors are much
better localized in the memory. If a cache line contains data not needed for
the current site it is most probably some data of a neighboring site, which
would be needed soon. Using the original structure this additional data of the
next bytes of the ``site major'' structure is another physical variable, which
is most likely not needed by the next calculational steps. Thus, ``site
major'' structure effectively spoils the memory bandwidth, which is one of the
most important factors of the performance. Similarly to Ref. \cite{G01}
by changing the ``site major'' code to a ``field major'' one  
we observed a speedup factor of $\approx$2 in the performances.

\begin{figure}[t]\begin{center}
\epsfig{file=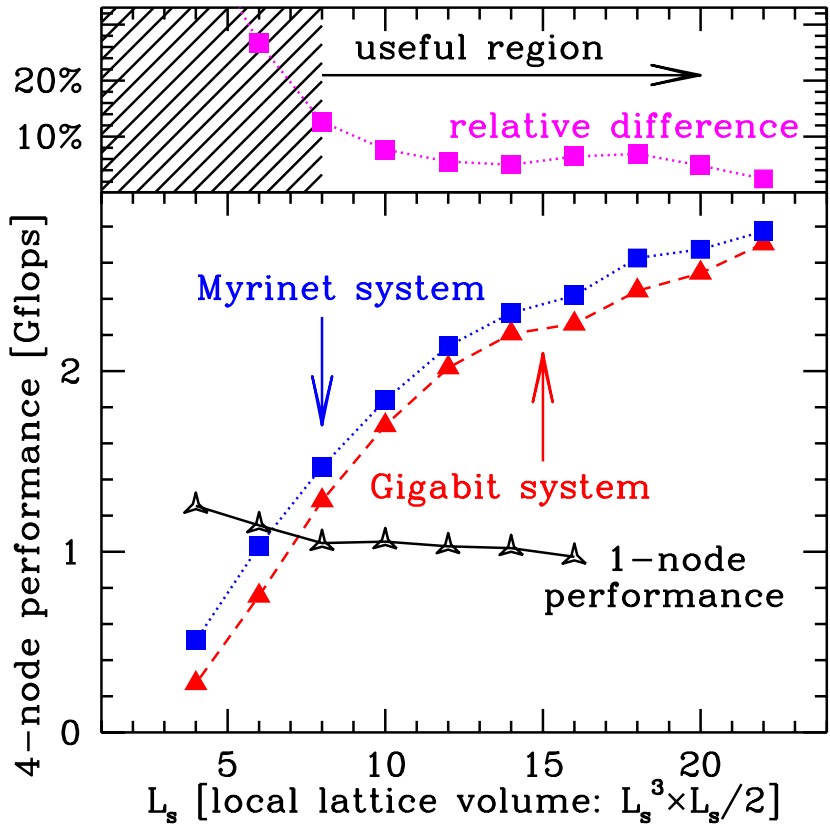,height=11.5cm}
\caption[hl]{
Sustained staggered performance of a 4-node system as a 
function of the local lattice
volume V=$L_s^3\times L_s/2$ (lower panel). The 
node topology is 2$\times$2, thus the total lattice volume is 
$L_s^3\times 2L_s$. The two almost parallel lines 
represent the results obtained by a myrinet system (upper 
line; these tests were perfomed on the myrinet
cluster of the DESY Theory group, Hamburg) and by  
our gigabit communication (lower line).
The one-node performance is also shown by the almost
horizontal line with stars. Clearly, communication is of no use
if the one-node performance is better than the 4-node one.
Thus, the useful region is given by local lattices
$8^3\times 4$ or larger. The relative difference
between the performances of the myrinet and gigabit systems
is shown by the upper panel. As it can be seen in the
useful region the relative difference is always less than 12\%.
\label{myri_gb}}
\end{center}\end{figure}

The third ingredient of our improvement was a better cache management.  
The number of successful cache-hits was enhanced by extensive use of the
``prefetch'' instruction. This technique further increased the sustained
performance of the code.  

One of the most obvious features of Figure~\ref{perf} is the sharp drop of the
performance when one turns on the communication. The most economic solution
is to turn on the communication only if the memory is insufficient for the
single-node mode. Smaller lattices can be also studied by using the 
communication (for instance for thermalization or parameter tuning); 
however, in these cases the communication overhead increases somewhat. E.g. 
having twice as large surface/volume ratio than the optimal one 
on a fixed number of communicating nodes we observed 
an additional 5-10\% reduction of the sustained performance.

It is instructive to study
the question, how the communication overhead can be reduced. 
We needed more than 1 node
when lattices of 18$^3\cdot$36 or larger were studied. With a somewhat better
memory allocation lattices of 20$^3\cdot$40 can still fit in 1 node. In this case 
the volume/surface ratio of the local sub-lattice gets better resulting in a
better communication overhead (e.g. in this case one would
 obtain 30\% instead of 35\%, which is slightly better).
The typical measured communication bandwidth in QCD applications was 
around 400~Mbit/s. This is far less than the maximal theoretical bandwidth of
1000 Mbit/s. Since we are using the most straightforward communication based
on TCP/IP it is feasible that a more effective driver can reach a bandwidth
around 800~Mbit/sec. In this case the communication overhead would be around
20\%. 

There is a simple theoretical consideration\footnote{we thank 
our referee for this transparent interpretation} to describe the 
performance of our architecture as $P(V,A)=P_0(V)/(1+c\cdot A/V)$,
where $P_0$ is the performance without communications for a given
local lattice volume $V$ on each node. $A$ is the number of
``surface'' lattice sites (per node) which participate in the
communication. If each node holds $L_x\cdot L_y$ lattice sites
in the communication plane one has $A/V=2(L_x+L_y)/(L_x\cdot L_y)$,
which becomes $A/V=4/\sqrt{N_p}/L_s$ if the $L_s^3\cdot 2L_s$ 
lattice can be distributed in an optimal way ($L_x=L_y$) over the
2-dimensional array of $N_p$ processors. Considering only a 
limited range of local lattice volumes the variation of $P_0(V)$ due 
to cache effects is negligible. From the points for $L_s$=12,...,16
in Figure~\ref{perf} one can read off a value 
$P_0\approx$ 1.63~Gflops. The above formula for $P(V,A)$
reproduces all measured values shown in Figure~\ref{perf}
for $L_s\ge$12 with an accuracy of a few percent using c=1.55.

Note, that the
"simulation scale" of a given physics problem depends on the
lattice size used plus the CPU cost to obtain a certain result. The
communication overhead of the present architecture is relatively
large, i.e. even in the optimal case (the local lattice is large,
occupying the whole memory) it is 30--40\%. Taking smaller local
lattices the sustained performance further drops. According to
Figure~\ref{myri_gb} taking local volumes, which are 80 times smaller than the
maximal ones results in an additional 50\% drop in the sustained
performance. When one uses the system as a scalable one
and simulates relatively small local lattices this additional
50\% reduction should be also taken into account.

\section{Conclusion}

We have presented a description of the status of our PC-based
parallel computer project for lattice QCD. 
Nearest-neighbor communication 
is implemented in a two-dimensional mesh. Communication goes through 
Gigabit Ethernet cards. 
Some details of the communication software have been given. 
Presently, we work on a better driver to increase the communication
bandwidth (in this way we expect to half the communication overhead).
As it is usual for many PC clusters for our system it is also advisable to
put sub-lattices on a relatively small number of nodes. In
this case one simulates several independent smaller lattices in parallel. 
 Using
this technique one reaches an optimum for the surface to volume ratio
and minimizes the time fraction for communication.
For larger lattices the whole system should be used. 

The saturated dynamical QCD sustained performance of a 128 node system 
with communication is around 70~Gflops for staggered quarks and 
110~Gflops for Wilson quarks (these numbers can be seen as a sort 
of average performances for typical applications using subsystems 
with different node numbers and relatively large local lattices; 
see Figures \ref{perf},\ref{myri_gb}).

Finally, we would like to emphasize that the system presented here
\hfill\break
\hspace*{4mm} 1. is a scalable parallel computer with 
nearest-neighbor communication,\hfill\break
\hspace*{4mm} 2. can be assembled using standard PC components,\hfill\break
\hspace*{4mm} 3. takes advantage of the free software environment,\hfill\break
\hspace*{4mm} 4. has a very good price/sustained-performance ratio for full 
lattice  QCD (with Wilson quarks\\
\hspace*{9mm} less than \$1/Mflops, for staggered quarks \$1.5/Mflops),\hfill\break
\hspace*{4mm} 5. due to its large memory and its large total disk bandwidth 
it can be used for heavy quark \\
\hspace*{9mm} phenomenology.

\section{Acknowledgment}
We thank F.~Csikor and Z.~Horv\'ath for their continuous help. We thank them 
and 
J.~Kuti, 
Th.~Lippert, 
I.~Montvay, 
K.~Schilling, 
H.~Simma and 
R.~Tripiccione 
for suggestions and careful
reading of the manuscript. The benchmark tests were done by modifying
the MILC Collaboration's public code (see http://physics.indiana.edu/\~{
}sg/milc.html).
This work was supported by Hungarian Science Foundation Grants under Contract
Nos. OTKA\-T37615\-T34980/\-T29803/\-T22929\-M37071/\-OM-MU-708/\-IKTA111/\-FKFP220/00.

\newpage
\appendix
\section[Appendix]{Example source code}
The following source code demonstrates the usage of sse instructions and
communication.

\begin{verbatim}
/**************************************************************
 *                                                            *
 *                     File: example.c                        *
 *   Get SU(3) vectors from neighbors, multiply them by the   *
 *              local links and add the result                *
 *                                                            *
 **************************************************************/

#include<stdio.h>
#include<stdlib.h>
#include"gb_lib.h"

#define DIR_RIGHT 0
#define DIR_UP    1
#define DIR_DOWN  2
#define DIR_LEFT  3

/* su3 vector conveniently aligned for sse operations a_r[3] and a_i[3]
   are used only to maintain alignment */
typedef struct{
    float a_r[4];
    float a_i[4];
} sse_su3_vector __attribute__ ((aligned (16)));

/* su3 matrix */
typedef struct{
    sse_su3_vector e[3];
} sse_su3_matrix __attribute__ ((aligned (16)));

void sse_mult_su3_mat_vec(sse_su3_matrix *u, sse_su3_vector *src,
                          sse_su3_vector *dest);
void sse_add_su3_vec(sse_su3_vector *v1, sse_su3_vector *v2,
                     sse_su3_vector *dest);

void init_links(sse_su3_matrix *link){
    /* Link initialization code may come here */
}

void init_vec(sse_su3_vector *link){
    /* Vector initialization code may come here */
}

int main(int argc, char *argv[]){

    static sse_su3_vector vect     __attribute__ ((aligned (16)));
    static sse_su3_vector temp1    __attribute__ ((aligned (16)));
    static sse_su3_vector temp2    __attribute__ ((aligned (16)));
    static sse_su3_vector temp3    __attribute__ ((aligned (16)));
    static sse_su3_vector temp4    __attribute__ ((aligned (16)));
    static sse_su3_vector res      __attribute__ ((aligned (16)));
    static sse_su3_matrix link[2]  __attribute__ ((aligned (16)));

    int nodes_x,nodes_y;

    /* Open communication channels */
    if (gb_open(&nodes_x,&nodes_y)!=0) {
        printf("Unable to open Gigabit communication channels\n");
        exit(-1);
    }
    printf("Machine topology: %d x %d nodes\n",nodes_x,nodes_y);
    /* Initialize links and vector */
    init_links(link);
    init_vec(&vect);

    /* Get vectors from neighbors */
    gb_send_recv(DIR_LEFT,(char *)&vect,(char *)&temp1,
                 sizeof(sse_su3_vector));
    gb_send_recv(DIR_DOWN,(char *)&vect,(char *)&temp2,
                 sizeof(sse_su3_vector));

    /* Perform multiplication with links */
    sse_mult_su3_mat_vec(&(link[0]),&temp1,&temp3);
    sse_mult_su3_mat_vec(&(link[1]),&temp2,&temp4);
    sse_add_su3_vec(&temp3,&temp4,&res);
    gb_close();
    return(0);
}

;*********************************************************
;                                                        *
;                    File su3_sse.asm                    *
; SU(3) matrix-vector multiplication and vector addition *
;                Compile with nasm v0.98                 *
;                                                        *
;*********************************************************


        BITS 32
        GLOBAL sse_mult_su3_mat_vec
        GLOBAL sse_add_su3_vec

REAL    equ 0
IMAG    equ 16

C0      equ 0
C1      equ 32
C2      equ 64

; SU(3) matrix-vector multiplication
; C prototype:
; 
; void sse_mult_su3_mat_vec(sse_su3_matrix *u, sse_su3_vector *src, 
;                           sse_su3_vector *dest);
; dest=u*src

sse_mult_su3_mat_vec:
        mov eax,[esp+04h]
        mov ecx,[esp+08h]
        mov edx,[esp+0ch]
        prefetcht0 [eax]
        movaps xmm6,[ecx+REAL]     ; first element of the vector
        movaps xmm7,[ecx+IMAG]
        shufps xmm6,xmm6,00000000b ; real part in xmm6
        shufps xmm7,xmm7,00000000b ; imaginary part in xmm7
        movaps xmm4,[eax+C0+REAL]  ; first column of the matrix
        movaps xmm5,[eax+C0+IMAG]  ; to (xmm4,xmm5)
        movaps xmm0,xmm4           ; complex multiplication
        mulps xmm0,xmm6
        movaps xmm2,xmm7
        mulps xmm2,xmm5
        subps xmm0,xmm2            ; real part goes to xmm0
        movaps xmm1,xmm4
        mulps xmm1,xmm7
        mulps xmm5,xmm6
        addps xmm1,xmm5            ; imaginary part goes to xmm1
        movaps xmm6,[ecx+REAL]     ; second element of the vector  
        movaps xmm7,[ecx+IMAG]
        shufps xmm6,xmm6,01010101b
        shufps xmm7,xmm7,01010101b
        movaps xmm4,[eax+C1+REAL]  ; second column of the matrix
        movaps xmm5,[eax+C1+IMAG]
        movaps xmm2,xmm4
        mulps xmm2,xmm6
        addps xmm0,xmm2
        movaps xmm2,xmm7
        mulps xmm2,xmm5
        subps xmm0,xmm2            ; real part added to xmm0
        mulps xmm4,xmm7
        addps xmm1,xmm4
        mulps xmm5,xmm6
        addps xmm1,xmm5            ; imaginary part added to xmm1
        movaps xmm6,[ecx+REAL]     ; third element of the vector  
        movaps xmm7,[ecx+IMAG]
        shufps xmm6,xmm6,10101010b
        shufps xmm7,xmm7,10101010b
        movaps xmm4,[eax+C2+REAL]  ; third column of the matrix
        movaps xmm5,[eax+C2+IMAG]
        movaps xmm2,xmm4
        mulps xmm2,xmm6
        addps xmm0,xmm2
        movaps xmm2,xmm7
        mulps xmm2,xmm5
        subps xmm0,xmm2            ; real part added to xmm0
        mulps xmm4,xmm7
        addps xmm1,xmm4
        mulps xmm5,xmm6
        addps xmm1,xmm5            ; imaginary part added to xmm1
        movaps [edx+REAL],xmm0     ; store the result to dest
        movaps [edx+IMAG],xmm1
        ret

; SU(3) vector addition
; C prototype:
; 
; void sse_add_su3_vec(sse_su3_vector *v1, sse_su3_vector *v2, 
;                      sse_su3_vector *dest);
; dest=v1+v2

sse_add_su3_vec:        
        mov eax,[esp+04h]
        mov ecx,[esp+08h]
        mov edx,[esp+0ch]
        movaps xmm0,[eax+REAL]     ; real parts
        movaps xmm1,[eax+IMAG]     ; imaginary parts
        addps xmm0,[ecx+REAL]
        addps xmm1,[ecx+IMAG]
        movaps [edx+REAL],xmm0        
        movaps [edx+IMAG],xmm1        
        ret
\end{verbatim}

\end{document}